\documentclass[article]{IEEEtran}

\usepackage[T1]{fontenc}
\usepackage{graphicx}
\usepackage{cite}
\usepackage{subcaption}

\usepackage{lipsum}
\usepackage{overpic}
\usepackage{amssymb}
\usepackage{amsmath}
\usepackage{amsthm}
\usepackage{array}
\usepackage[monochrome]{xcolor}
\usepackage{url}

\IEEEoverridecommandlockouts

\theoremstyle{plain}

\newtheorem{assumption}{Assumption}
\newtheorem{corollary}{Corollary}
\newtheorem{remark}{Remark}
\newtheorem{proposition}{Proposition}

\newcommand{\vect}[1]{\mathbf{#1}}

\def\diag{\mathrm{diag}}

\def\tr{\mathrm{tr}}

\def\Htran{\mbox{\tiny $\mathrm{H}$}}
\def\Ttran{\mbox{\tiny $\mathrm{T}$}}
\def\CN{\mathcal{N}_{\mathbb{C}}} %Complex Gaussian
\def\imagunit{\mathsf{j}} % Imaginary number

\def\sinc{\mathrm{sinc}}

\begin{document}

\title{\huge Electromagnetic Interference in RIS-Aided Communications}

\author{\IEEEauthorblockN{Andrea de Jesus Torres, \emph{Graduate Student Member, IEEE}, Luca Sanguinetti, \emph{Senior Member, IEEE}, Emil Bj{\"o}rnson, \emph{Senior Member, IEEE} \vspace{-0.7cm}
\thanks{
\indent This work was supported by the FFL18-0277 grant from the Swedish Foundation for Strategic Research and the Italian Ministry of Education and Research in the framework of the CrossLab project.\newline
\indent A. de Jesus Torres and L.~Sanguinetti are with the University of Pisa, Dipartimento di Ingegneria dell'Informazione, 56122 Pisa, Italy (luca.sanguinetti@unipi.it). E.~Bj\"ornson is with the Department of Computer Science, KTH Royal Institute of Technology, 10044 Stockholm, Sweden, and Department of Electrical Engineering, Link\"{o}ping University, 58183 Link\"{o}ping, Sweden (emilbjo@kth.se).}
% make the title area
}}
\maketitle

\begin{abstract}
The prospects of using a reconfigurable intelligent surface (RIS) to aid wireless communication systems have recently received much attention. Among the different use cases, the most popular one is where each element of the RIS scatters the incoming signal with a controllable phase-shift, without increasing its power. In prior literature, this setup has been analyzed by neglecting the electromagnetic interference, consisting of the inevitable incoming waves from external sources. In this letter, we provide a physically meaningful model for the electromagnetic interference that can be used as a baseline when evaluating RIS-aided communications. The model is used to show that electromagnetic interference has a non-negligible impact on communication performance, especially when the size of the RIS grows large. When the direct link is present (though with a relatively weak gain), the RIS can even reduce the communication performance. Importantly, it turns out that the SNR grows quadratically with the number of RIS elements only when the spatial correlation matrix of the electromagnetic interference is asymptotically orthogonal to that of the \textcolor{red}{effective} channel \textcolor{red}{(including RIS phase-shifts)} towards the intended receiver. Otherwise, the SNR only increases linearly.%
\end{abstract}


\vspace{-2mm}

\IEEEpeerreviewmaketitle

\section{Introduction}
Reconfigurable intelligent surface (RIS) is an umbrella term used for a two-dimensional array of passive elements that will (diffusely) reflect incident electromagnetic waves after ``passive'' analog filtering~\cite{Renzo2020b}. Each element filters the signal by potentially reducing the amplitude, incurring time delays, and/or changing the polarization~\cite{bjornson2021reconfigurable}. A basic use case of the RIS technology is illustrated in Fig.~\ref{figure_geometric_setup}, where an RIS is deployed to capture signal energy from the source proportional to its area and re-radiate it in the shape of a beam towards the intended receiver. Since the RIS is not amplifying the signal, a large surface area is typically required to achieve a given signal-to-noise ratio (SNR) at the receiver. 

 RIS-aided communications is an emerging topic that is receiving a lot of attention~\cite{Renzo2020b,bjornson2021reconfigurable} and several papers have identified potential benefits in terms of spectral efficiency~\cite{Qingqing2019} and energy efficiency~\cite{Huang2018a}. However, a common practice is to only consider the signals generated by the system and thereby neglecting the electromagnetic interference (EMI) or ``noise'' (or ``pollution'') that is inevitably present in any environment~\cite{Middleton-1977,Loyka-2004}. The EMI may arise from a variety of natural, intentional or non-intentional causes; for example, man-made devices and natural background radiation. Largely speaking, any \emph{uncontrollable} wireless signal produces EMI. Despite existing in any wireless communication system, it may have a severe effect in the setup of Fig.~\ref{figure_geometric_setup} since the EMI that impinges on the RIS, from arbitrary spatial directions, is captured with an energy that is proportional to its area and then re-radiated.
 While the direction of the re-radiated EMI might not be focused at the intended receiver, a significant portion (due to the large surface area of the RIS) of its energy might reach it and, thus, degrade the end-to-end SNR of the system, which is typically designed for optimal performance in presence of only thermal noise. 

In this letter, we present a physically meaningful model for the EMI that is produced by uncontrollable sources in the far-field of the RIS. The model is valid for arbitrary scattering and can be used as a baseline for the analysis and design of RIS-aided communications. Particularly, it is used to show that in  a random scattering environment, the EMI may have a severe impact on the SNR, especially when the size of the RIS grows large. Importantly, it turns out that the SNR grows linearly, not quadratically as in~\cite{Qingqing2019}, with the number of RIS elements. When a direct link with a non-negligible gain is present, the RIS can even reduce the communication performance. A heuristic method, based on the projected gradient descent method and knowledge of EMI statistics, is also proposed to optimize the RIS against both thermal noise and EMI. Numerical results show that slightly better performance can be achieved, but the scaling behavior of the SNR remains linear.

\emph{Reproducible Research}: Simulation code available at: \url{https://github.com/lucasanguinetti/EMI-RIS-Communications}

\begin{figure}[t!]
	\centering \vspace{-0.1cm}
	\begin{overpic}[width=.75\columnwidth,tics=10]{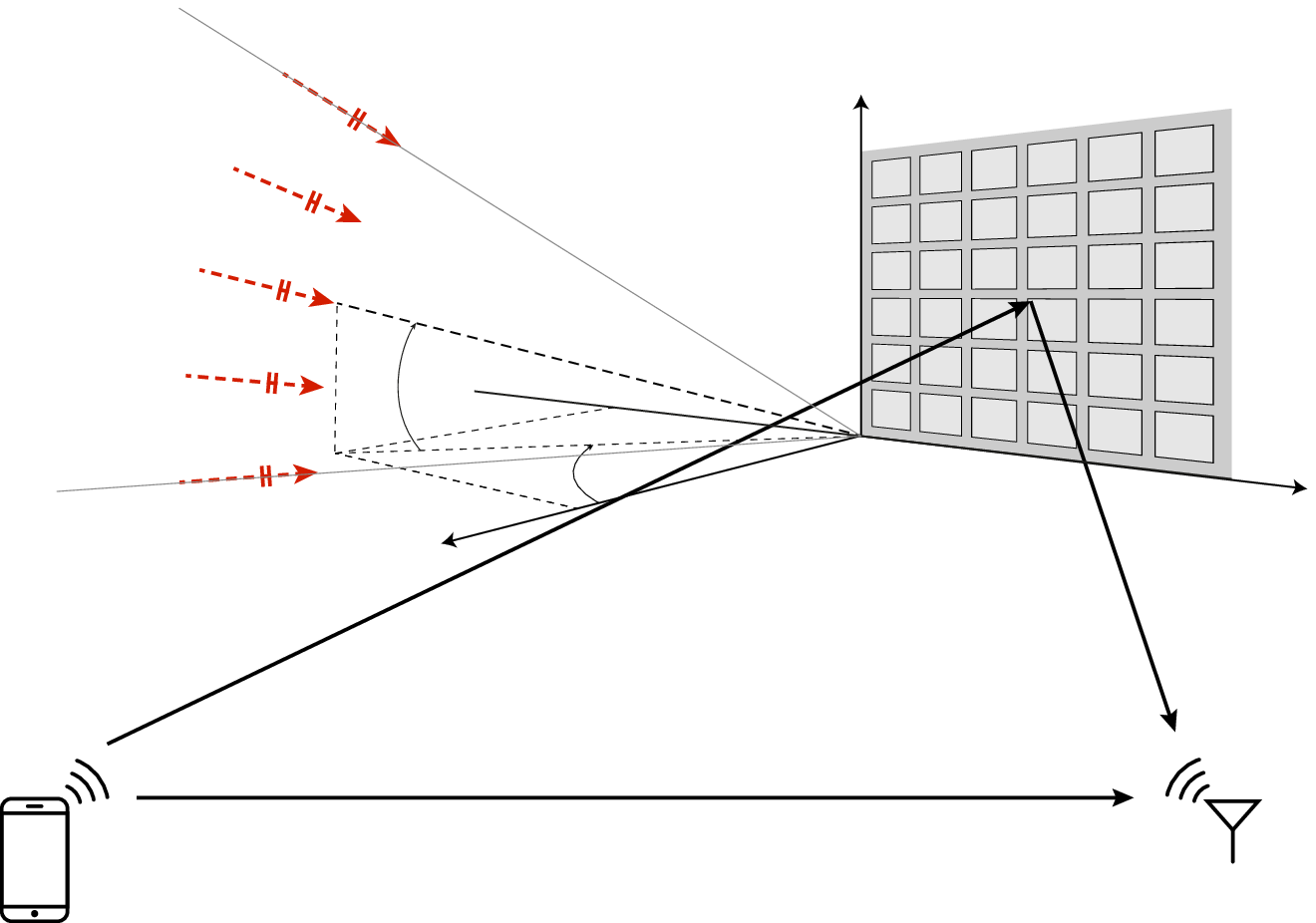}
		\put(10,1){\footnotesize Source}
		\put(21,25){\footnotesize ${\bf h}_1$}
		\put(88,26){\footnotesize ${\bf h}_2$}
		\put(47,12){\footnotesize ${h}_{\rm d}$}
		\put(27,68){\footnotesize \textcolor{red}{Electromagnetic}}
		\put(30,63){\footnotesize \textcolor{red}{interference}}
		\put(42,5){\footnotesize direct link}
		\put(80,1){\footnotesize Destination}
		\put(75,62){\footnotesize RIS}
		
		\put(33,30.3){\footnotesize ${\bf \hat{x}}$}
		\put(98,35){\footnotesize ${\bf \hat{y}}$}
		\put(67,62){\footnotesize ${\bf \hat{z}}$}
		
		\put(28,40){\footnotesize ${\bf {\theta}}$}
		\put(40.8,33.5){\footnotesize ${\bf {\varphi}}$}
		
\end{overpic} 
	\caption{RIS-aided communication system.}\vspace{-0.6cm}
	\label{figure_geometric_setup}  
\end{figure}

\vspace{-0.3cm}
\section{System Model}

Consider a single-antenna transmitter communicating wirelessly with a single-antenna receiver in a scattering environment, while being aided by an RIS equipped with $N$ reconfigurable elements. The $N$ RIS elements are deployed edge-to-edge on a two-dimensional square grid~\cite{Renzo2020b}. 
The setup is illustrated in Fig.~\ref{figure_geometric_setup} in a three-dimensional space, where a local spherical coordinate system is defined at the RIS with $\varphi$ being the azimuth angle and $\theta$ being the elevation angle. The elements have an area $A$ and are indexed row-by-row by $n=1,\ldots,N$. Hence, the $n$th element is located at $\vect{u}_n= [u_{x,n},u_{y,n}, 0]^{\Ttran}$
where $u_{x,n} = - \frac{(\sqrt{N}-1)\sqrt{A}}{2} + \sqrt{A} \mod(n-1,\sqrt{N})$ and $u_{y,n} = \frac{(\sqrt{N}-1)\sqrt{A}}{2} - \sqrt{A} \left\lfloor \frac{n-1}{\sqrt{N}} \right\rfloor$. The channel vector between the source and RIS is $\vect{h}_1= [h_{1,1}, \, \ldots, \, h_{1,N}]^{\Ttran}$  and the channel vector between the RIS and receiver is $\vect{h}_2 = [h_{2,1}, \, \ldots, \, h_{2,N}]^{\Ttran}$. The RIS configuration is determined by the diagonal matrix $\boldsymbol{\Theta} =\boldsymbol{\Gamma}\boldsymbol{\Phi}$ with $\boldsymbol{\Gamma} = \diag(\gamma_1,\ldots,\gamma_N)$ and $\boldsymbol{\Phi} =  \diag(e^{\imagunit \phi_1},\ldots,e^{\imagunit \phi_N})$. Here, $\gamma_1,\ldots,\gamma_N \in(0,1]$ are the amplitude scattering variables (describing the fraction of the incident signal power that is scattered) and $\phi_1,\ldots,\phi_N\in[0,2\pi)$ are the phase-shift variables (describing the delays of the scattered signals). 
\vspace{-0.3cm}
\subsection{Signal Model}
The received signal ${\bf x} \in \mathbb{C}^N$ at the RIS is
\begin{equation} \label{eq:received-signal1}
	{\bf x} =  \vect{h}_1 s + {\bf n}
\end{equation}
where $s$ is the transmitted symbol with power $P=\mathbb{E}\{ |s|^2\}$ and ${\bf n}\in \mathbb{C}^N$ is the EMI, produced by the incoming, uncontrollable electromagnetic waves.
The received signal $y \in \mathbb{C}$ at the destination in Fig.~\ref{figure_geometric_setup} is
\begin{align}
	y =  \vect{g}_2^{\Htran} {\bf x} + h_{\mathrm{d}} s + w \label{eq:received-signal2}
\end{align}
where $\vect{g}_2 = \boldsymbol{\Theta} \vect{h}_2$ is the \textcolor{red}{\emph{effective}} channel \textcolor{red}{(including RIS phase-shifts)} and $h_{\mathrm{d}} \in \mathbb{C}$ is the channel gain of the direct path. The noise $w\sim\CN(0,\sigma^2_w)$ accounts for any uncontrollable factor (e.g., of electromagnetic or hardware nature) disturbing the signal reception, except for the EMI scattered by the RIS, because its statistics depend on the RIS properties. Plugging~\eqref{eq:received-signal1} into~\eqref{eq:received-signal2} yields \begin{align}
	y=\left(\vect{g}_2^{\Htran}\vect{h}_1 + h_{\mathrm{d}}\right) s+ \vect{g}_2^{\Htran}{\bf n} + w.\label{eq:received-signal3}
\end{align}
We want to evaluate the impact of the EMI (or electromagnetic noise) ${\bf n}$ on communication performance, which is neglected in the RIS literature. A statistical model is provided next. 

\vspace{-0.2cm}
\subsection{Electromagnetic Interference Modeling}
The EMI ${\bf n}$ is produced by a superposition of continuum of incoming plane waves that are generated by external sources.\footnote{Note that even a single spherical wave can be expanded as a continuum of plane waves.}
Suppose the waves are generated in the far-field of the half-space in front of the RIS. Each one can thus be modeled as a plane wave that reaches the RIS from a particular azimuth angle $\varphi \in [-\pi/2,\pi/2)$ and elevation angle $\theta\in [-\pi/2,\pi/2)$; see Fig.~\ref{figure_geometric_setup}.
The EMI  field ${\bf n}$ is thus (e.g.,~\cite{Wallace-2008})
\begin{align}\label{eq:noise_r}
{\bf n} = \iint_{-\pi/2}^{\pi/2} \!\!{\boldsymbol \nu}( \varphi, \theta)  d \varphi d \theta
\end{align}
where $ {\boldsymbol \nu}( \varphi, \theta)\in \mathbb{C}^N$ has entries
\begin{align}\label{eq:noise_angle}
 {\boldsymbol \nu}( \varphi, \theta) = {a}(\varphi,\theta)e^{\imagunit \vect{k}(\varphi, \theta)^{\Ttran}\vect{u}_n}.
\end{align}
In~\eqref{eq:noise_angle}, $\vect{k}(\varphi, \theta)= \frac{2\pi}{\lambda}[\cos(\theta) \cos(\varphi), \cos(\theta) \sin(\varphi),\sin(\theta)]^{\Ttran}$ is the wave vector
that describes the phase variation of the plane wave with respect to the three Cartesian coordinates at the receiving volume, and ${a}(\varphi,\theta)$ is a zero-mean, complex-Gaussian random process with 
\begin{align}\label{eq:noise_coefficients}
\mathbb{E}\{{a}(\varphi,\theta){a}(\varphi^\prime, \theta^\prime)\} =A {\sigma}^2f(\varphi,\theta) \delta(\varphi-\varphi^\prime)\delta(\theta-\theta^\prime)
\end{align}
where ${\sigma}^2f(\varphi,\theta)$ is the angular density of interference power with $\iint_{-\pi/2}^{\pi/2} f(\varphi,\theta) d \varphi d \theta=1$ and ${\sigma}^2$ is measured in W/m$^2$. 
From~\eqref{eq:noise_r}, using~\eqref{eq:noise_angle} and \eqref{eq:noise_coefficients} we obtain $
\mathbb{E}\{{\bf n}{\bf n}^{\Htran}\} = A{\sigma}^2 \vect{R}$
with
\begin{align}
[\vect{R}]_{n,m} = \iint_{-\pi/2}^{\pi/2} e^{\imagunit\vect{k}(\varphi,\theta)^{\Ttran}(\vect{u}_n-\vect{u}_m)} f(\varphi,\theta) d \varphi d \theta . \label{eq:R1_expression}
\end{align}
The following model is thus valid for any arbitrary $f(\varphi,\theta)$. \begin{corollary} \label{cor:Rayleigh-distributions}
The EMI $\vect{n}$ is distributed as $\vect{n} \sim \CN(\vect{0},A\sigma^2\vect{R}) $ where the $(n,m)$th element of $\vect{R}$ is given by \eqref{eq:R1_expression}.
\end{corollary}
Notice that $\frac{1}{N}\tr (\vect{R}) = 1$. The nature of EMI makes it reasonable to assume that the electromagnetic waves are impinging from directions spanning a large angular interval. With uniform distribution from all angles (i.e., isotropic conditions),~\eqref{eq:R1_expression} reduces to \cite[Prop. 1]{Bjornson21}
\begin{align}\label{eq:correlation_matrix}
[\vect{R}^{\textcolor{red}{\rm{iso}}}]_{n,m} =\sinc \left(\frac{2||\vect{u}_n-\vect{u}_m||}{\lambda}\right)
\end{align}
where $\| \cdot \|$ denotes the Euclidean norm.

\begin{remark}
In wireless communications, the noise samples are typically modeled as independent circularly-symmetric Gaussian random variables, i.e., $\vect{R} = {\bf I}_N$. This is never the case for the noise (interference) samples of electromagnetic nature. Even in the special case of an isotropic angular distribution,~\eqref{eq:correlation_matrix} shows that $\vect{R} = {\bf I}_N$ only when $||\vect{u}_n-\vect{u}_m|| = i \lambda/2$,  with $i \in \mathbb{Z}$, i.e., all the RIS elements are positioned along a straight line at a spacing of an integer multiple of $\lambda/2$. This can never happen with a two-dimensional RIS.
%This never happens in a planar surface.
\end{remark}
\vspace{-0.5cm}

\subsection{Channel Modeling}
The channels $\{{\bf h}_1$, ${\bf h}_2$, $h_{\mathrm{d}}\}$ can be deterministic or stochastic depending on propagation conditions. 
With random scattering, $\vect{h}_1,\vect{h}_2$ are independent and distributed as~\cite[Cor. 1]{Bjornson21}\,\, 
\begin{equation}\label{eq:channel_distribution}
\vect{h}_i \sim \CN(\vect{0},A\beta_i\vect{R}_i) \quad i=1,2
\end{equation}
where $\beta_i$ for $i=1,2$ is the average attenuation intensity and the $(n,m)$th element of $\vect{R}_i$ is given by \eqref{eq:R1_expression} with power angular density $f_i(\varphi,\theta)$. The direct path $h_{\mathrm{d}}$ has a Rayleigh fading distribution $h_{\mathrm{d}} \sim \CN(0, \beta_{\mathrm{d}})$, where $\beta_{\mathrm{d}}$ is the variance.

\section{Performance analysis}
Assume that perfect channel knowledge is available in the system. From~\eqref{eq:received-signal3}, the effective SNR at the destination is\footnote{We deliberately use SNR, instead of SINR (signal-to-interference-plus-noise ratio), to stress that EMI must be thought of as noise in the wireless communications parlance since it is generated by uncontrollable signals.}
\begin{align}\label{eq:SNR}
{\mathrm {SNR }} &= \frac{P |\vect{g}_2^{\Htran} \vect{h}_1 + h_{\mathrm{d}}|^2}{A \sigma^2 \vect{g}_2^{\Htran} \vect{R}\vect{g}_2 + \sigma_w^2}.
\end{align}
With the optimal phase-configuration against thermal noise, e.g.,~\cite{Huang2018a}, $\phi_n = \arg(h_{1n}\textcolor{red}{h_{2n}^*})-\arg(h_{\mathrm{d}})$ and~\eqref{eq:SNR} becomes
\begin{equation}\label{eq:optimized_SNR}
{\mathrm {SNR}} = 
\frac{1}{ \frac{A \sigma^2}{\sigma_w^2} \vect{g}_2^{\Htran} \vect{R}\vect{g}_2 + 1} {\overline {\mathrm {SNR}}}
\end{equation}
where 
\begin{equation}\label{eq:optimized_SNR_noNoise}
{\overline {\mathrm {SNR}}} =\frac{P}{\sigma_w^2} \left(\sum_{n=1}^N \gamma_n|h_{1n} h_{2n}|  + | h_{\mathrm{d}}| \right)^2
\end{equation}
is the maximized SNR in absence of EMI \cite{Huang2018a}. The impact of the EMI on the optimized SNR in~\eqref{eq:optimized_SNR} is next quantified. This allows us to make comparisons with the scaling behaviours reported in previous literature that neglected EMI. Notice that the optimized SNR plays a key role in the computation of other metrics such as the ergodic capacity and outage probability. {To proceed further, we define
\begin{equation}\label{eq:rho}
\rho = \frac{P\beta_1}{\sigma^2}
\end{equation}
which corresponds to the ratio between the received signal power and EMI power at each antenna element of the RIS.} 
\subsection{Without the Direct Link}
We start by analyzing the case without the direct link, i.e., $\beta_{\rm d} = 0$\,($-\infty$\,dB). We consider a system that uses a bandwidth $B=1$\,MHz and a transmit power $P = 23$\,dBm while the power spectral density of the thermal noise is $N_0 = -174$\,dBm/Hz. Consequently, $\sigma^2_w = N_0B = -114$\,dBm and $P/\sigma^2_w = 137$\,dB. We assume that $\lambda = 0.1$\,m and $A = (\lambda/4)^2 = 6.25\times10^{-4}$ m$^2$. The channels ${\bf h}_1$ and ${\bf h}_2$ are isotropic with \textcolor{red}{${\bf R}_1={\bf R}_2={\bf R}^{\rm{iso}}$}, and $A\beta_1 = -80$\,dB and $A\beta_2 = -70$\,dB. The EMI is also modeled as isotropic, i.e., \textcolor{red}{${\bf R}={\bf R}^{\rm{iso}}$}. We assume $\gamma_1=\ldots=\gamma_N=1$. Fig.~\ref{simulation-NoDirectPath} shows the average SNR in~\eqref{eq:optimized_SNR} as a function of $N$ when $\rho = 30$, $20$, and $10$\,dB which from~\eqref{eq:rho} correspond to $A\sigma^2 = -87$, $-77$, and $-67$\,dBm, respectively. Results are averaged over $1000$ channel realizations. The impact of EMI depends on its strenght compared to the $1$-term in the denominator of~\eqref{eq:optimized_SNR}. The impact is small when it is smaller than one. In the investigated scenarios, we see that the EMI has a non-negligible effect on the SNR already when $\rho = 30$\,dB; that is, the EMI power is $30$\,dB lower than the received signal power. The gap increases as $\rho$ decreases but also when $N$ increases. This is an undesirable effect since a physically large RIS is needed to compensate for propagation losses and increase $ {\overline {\mathrm {SNR}}}$ in~\eqref{eq:optimized_SNR_noNoise}. 

\begin{figure} 
        \centering\vspace{-5mm}
	\begin{overpic}[width=1\columnwidth,tics=10]{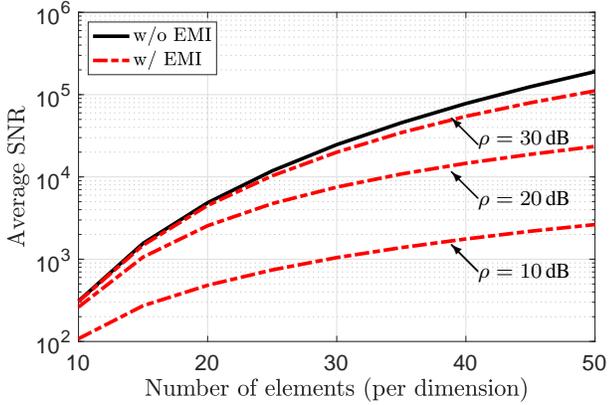}
	\put(73,39){\vector(-1,1){4}}
%	\put(70,41){\footnotesize $\sigma^2 = -70$\,dB}
		\put(73,39){\footnotesize $\rho = 30$\,dB}
	\put(73,31){\vector(-1,1){4}}
%	\put(70,33){\footnotesize $\sigma^2 = -60$\,dB}
		\put(73,30){\footnotesize $\rho = 20$\,dB}
	\put(73,20){\vector(-1,1){4}}
%	\put(70,22){\footnotesize $\sigma^2 = -50$\,dB}
	\put(73,19){\footnotesize $\rho = 10$\,dB}
\end{overpic} 
                \label{simulation-NoDirectPath} \vspace{-0.3cm}
        \caption{Average SNR achieved with an optimized RIS without the direct path for varying $N$ with EMI. We assume isotropic scattering for ${\bf n}$ and $\{{\bf h}_i\}$, i.e., \textcolor{red}{${\bf R}={\bf R}_1={\bf R}_2={\bf R}^{\rm{iso}}$}. The SNR achieved without EMI is reported as a reference.}
                        \label{simulation-NoDirectPath} \vspace{-0.5cm}
\end{figure}

 \vspace{-0.3cm}
\subsection{With the Direct Link}
\begin{figure} 
        \centering\vspace{-5mm}
	\begin{overpic}[width=1\columnwidth,tics=10]{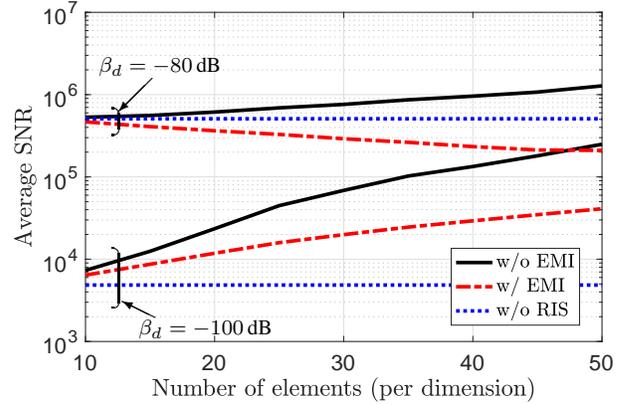}
	\put(15,50){\footnotesize $\beta_d = -80$\,dB}
	\put(23,49){\vector(-1,-1){5}}
	\put(17,42.5){\oval(2,4)[r]}
	
	\put(21,10.5){\footnotesize $\beta_d = -100$\,dB}
	\put(23.5,13){\vector(-2,1){5}}
	\put(17,19){\oval(2,9)[r]}
\end{overpic} \vspace{-0.3cm}
        \caption{Average SNR achieved with an optimized RIS for varying $N$ with the direct path and \textcolor{red}{${\bf R}={\bf R}_1={\bf R}_2={\bf R}^{\rm{iso}}$}. With EMI, we assume that {$\rho = 20$\,dB}.}
                        \label{simulation-WithDirectPath} \vspace{-0.45cm}
\end{figure}

We now consider a simulation scenario in which the direct path is present. Fig.~\ref{simulation-WithDirectPath} shows the optimized SNR in~\eqref{eq:optimized_SNR} for the same setup as in Fig.~\ref{simulation-NoDirectPath} with $\rho = 20$\,dB, but with $\beta_d= -100$\,dB and $-80$\,dB. The curves 'w/o RIS' refer to the SNR achieved without the aid of the RIS. We notice that the RIS can increase the SNR by orders-of-magnitude when the EMI is neglected, even in presence of a direct link. However, when it is considered, the gain reduces substantially when $\beta_d= -100$\,dB. Interestingly, the RIS even deteriorates the performance when the direct path gain is $\beta_d= -80$\,dB. The negative impact of having the RIS increases if $\beta_d$ grows. This is because when the direct link is present, a large number of elements $N$ is needed if the extra signal power delivered by the RIS should make a difference in the SNR calculation. However, the extra interference (or noise) that is reflected by the RIS can dominate over the thermal noise in the receiver even when the RIS is relatively small, thereby reducing the SNR. This makes it evident that the EMI plays a key role in RIS-aided communications.

 \vspace{-0.25cm}
\subsection{Power Scaling Law With Asymptotically Large RIS}
The results in Figs.~\ref{simulation-NoDirectPath} and~\ref{simulation-WithDirectPath} show that the scaling behavior of the SNR is much different with or without the EMI. Hence, we now study the asymptotic SNR~\eqref{eq:optimized_SNR} when $N\to \infty$. To this end, we recall the following result.
\begin{proposition}~\cite[Prop. 3]{Bjornson21} \label{prop:hardening}
In a rich non-isotropic scattering environment with $\vect{h}_1$ and $\vect{h}_2$ being independent and distributed as in~\eqref{eq:channel_distribution}, we have that, as $N \to \infty$,
\begin{equation} \label{eq:hardening}
\frac{\overline {\mathrm {SNR}}}{N^2} \overset{p}{\longrightarrow}  \frac{P}{\sigma_w^2}  \beta_1 \beta_2 \left(\frac{\pi}{4} A\right)^2
\end{equation}
where the convergence is in probability.
\end{proposition}
Proposition~\ref{prop:hardening} implies that the SNR in the absence of EMI scales quadratically as $N$ grows. This is an instance of the so-called ``square law'' property of RIS-aided communications, originally presented in~\cite{Wu2019a,Qingqing2020b} and commonly adopted in the RIS literature. To derive the power scaling law with EMI, \textcolor{red}{we denote $\mathbb{E}\{{\bf g}_2{\bf g}_2^{\Htran}\}= A\beta_2\overline {\bf R}$} and make the following technical assumptions:
\begin{assumption} \label{Assumption_1}$\lim \inf_N \frac{1}{N}\tr({\bf R}) > 0 ,\lim \sup_N ||{\bf R}||_2 < \infty $.
\end{assumption}
\begin{assumption} \label{Assumption_2}\textcolor{red}{$\lim \inf_N \frac{1}{N}\tr(\overline {\bf R}) > 0$, $\lim \sup_N ||\overline {\bf R}||_2 < \infty $.}
\end{assumption}

\begin{proposition} \label{prop:hardening_noise}
Assume a rich non-isotropic scattering environment with $\vect{h}_1$ and $\vect{h}_2$ being independent and distributed as in~\eqref{eq:channel_distribution}. If Assumptions~\ref{Assumption_1} and~\ref{Assumption_2} hold and  
\begin{equation}\label{eq:asymptotical_ortho}
 \lim \inf_N \frac{1}{N}\tr(\textcolor{red}{\overline {\bf R}}{\bf R}) > 0
\end{equation}
then, as $N \to \infty$,
\begin{equation} \label{eq:hardening_noise}
\frac{{\mathrm {SNR}}}{N} \overset{p}{\longrightarrow} \frac{P}{\sigma^2}  \frac{\beta_1}{\alpha}\left(\frac{\pi}{4}\right)^2
\end{equation}
where $\alpha = \lim_N\frac{1}{N}\tr(\textcolor{red}{\overline {\bf R}}\vect{R})$ and convergence is in probability.
\end{proposition}
\begin{IEEEproof}
After dividing both sides by $N$,~\eqref{eq:optimized_SNR} reduces to (multiplying and dividing the left-hand-side by $N$)
\begin{equation}\label{eq:hardening_noise_1}
\frac{{\mathrm {SNR}}}{N} = 
\frac{1}{\frac{A \sigma^2}{\sigma_w^2} \frac{1}{N}\vect{g}_2^{\Htran} \vect{R} \vect{g}_2 + \frac{1}{N}}\frac{\textcolor{red}{\overline {{\mathrm {SNR}}}}}{N^2}. 
\end{equation}
\textcolor{red}{Under Assumptions~\ref{Assumption_1} and~\ref{Assumption_2}, we observe that $\frac{1}{N}\vect{g}_2^{\Htran} \vect{R} \vect{g}_2$ converges to $A \beta_2 \alpha$ almost surely.}
By using the continuous mapping theorem~\cite{CouilletRMT}, the first term in~\eqref{eq:hardening_noise_1} converges almost surely to $\sigma_w^2/ (A\sigma^2 A\beta_2\alpha)$. Since almost sure convergence implies convergence in probability,~\eqref{eq:hardening_noise} follows since ${\mathrm {SNR}}/N^2$ converges to~\eqref{eq:hardening} in probability.
\end{IEEEproof}

\begin{figure} 
        \centering\vspace{-6mm}
	\begin{overpic}[width=0.95\columnwidth,tics=10]{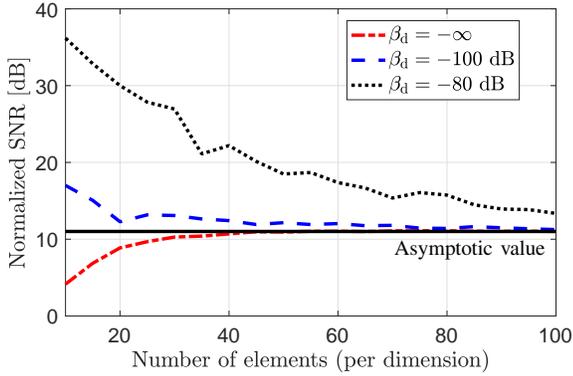}
	\put(65,19){\footnotesize {Asymptotic value}}
\end{overpic} 
\vspace{-2mm}
        \caption{Scaling behavior of the effective SNR, normalized to $N$, for different values of $\beta_{\rm d}$ and \textcolor{red}{${\bf R}={\bf R}_1={\bf R}_2={\bf R}^{\rm{iso}}$}. We assume {$\rho = 20$\,dB}.}
                        \label{simulation-Asymptotic-normalized} \vspace{-0.45cm}
\end{figure}

Proposition~\ref{prop:hardening_noise} shows that the SNR scales linearly, not quadratically, as $N$ grows. The signal power grows as $N^2$ but is divided by the EMI, which grows as $N$, plus the thermal noise that is independent of $N$. Technically speaking, the new scaling behavior is a direct consequence of the condition~\eqref{eq:asymptotical_ortho}, which states that $\textcolor{red}{\overline {\bf R}}$ and ${\bf R}$ are \emph{not asymptotically orthogonal}, as $N\to \infty$. This implies that the common subspace of $\textcolor{red}{\overline {\bf R}}$ and ${\bf R}$ has dimension and eigenvalues that grow linearly with $N$~\cite{Sanguinetti-2020}. When~\eqref{eq:asymptotical_ortho} does not hold true, that is, $\textcolor{red}{{\overline {\bf R}}}$ is asymptotically orthogonal to ${\bf R}$ (i.e., $\frac{1}{N}\tr(\textcolor{red}{\overline {\bf R}}\vect{R}) \to 0$ as $N\to \infty$), then the effect of the EMI vanishes and the SNR grows quadratically with $N\to \infty$. Although possible, this is unlikely to happen in practice since the EMI accounts for uncontrollable electromagnetic waves that can impinge on the RIS from arbitrary (and many) spatial directions. 
{\color{red}Since the RIS configuration matches  $\vect{g}_2$ to $\vect{h}_1$, the EMI has a lower impact when $\vect{R}$ and 
$\vect{R}_1$ have very different subspaces.}

\textcolor{red}{To validate Proposition~\ref{prop:hardening_noise}}, Fig.~\ref{simulation-Asymptotic-normalized} illustrates the behavior of ${{\mathrm {SNR}}}/{N}$ for varying $N$, under isotropic scattering conditions for both EMI and channels $\{{\bf h}_1,{\bf h}_2\}$. In agreement with Proposition~\ref{prop:hardening_noise}, ${{\mathrm {SNR}}}/{N}$ converges quickly to the limit in~\eqref{eq:hardening_noise}, irrespective of the direct link strength. The results show that, when there is a non-negligible direct link, the normalized SNR decreases as $N$ grows large, until it converges to the limit; that is, the SNR scales sub-linearly with $N$ before convergence. Hence, in Figs.~\ref{simulation-NoDirectPath} and~\ref{simulation-WithDirectPath}, the SNR reduction due to EMI only represents the gap for a given $N$, while the asymptotic gap grows without bound.

\begin{figure} 
        \centering\vspace{-6mm}
	\begin{overpic}[width=0.95\columnwidth,tics=10]{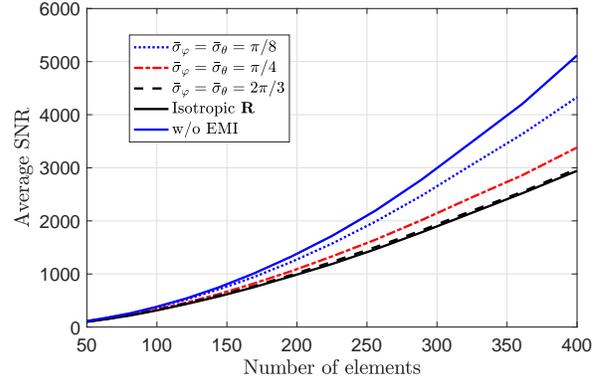}
\end{overpic} 
\vspace{-2mm}
        \caption{Scaling behavior of the average SNR for non-isotropic scattering for ${\bf n}$ and $\{{\bf h}_1,{\bf h}_2\}$ in the same setup of Fig.~\ref{simulation-NoDirectPath} with $\rho=20$\,dB. The case with isotropic EMI is also reported.}
                        \label{simulation-Non-Isotropic-Scattering} \vspace{-0.5cm}
\end{figure}

To quantify the impact of modeling of \textcolor{red}{${\bf R}_1,{\bf R}_2$ (and thus $\overline{\bf R}$)} and ${\bf R}$, we assume the power angular density of ${\bf h}_i$ for $i=1,2$ is 
\begin{align}\label{eq:non-isotropic_pdf}
f_i(\varphi,\theta)  = \frac{c}{2 \pi \bar\sigma_{i,\varphi} \bar\sigma_{i,\theta}} e^{-\frac{(\varphi - \bar\varphi_i)^2}{2 \bar\sigma_{i,\varphi}^2}}e^{-\frac{(\theta - \bar\theta_i)^2}{2 \bar\sigma_{i,\theta}^2}} \cos (\theta)
\end{align}
where $c$ is a constant such that $\iint_{-\pi/2}^{\pi/2} f_i(\varphi,\theta) d\varphi d \theta = 1$. 
This represents a concentration of plane waves around the nominal angle pair $(\bar\varphi_i,\bar\theta_i)$ with a Gaussian angular distribution.
We fix $\bar\varphi_1= 0$, $\bar\varphi_2 = \pi/4$, $\bar\theta_1=\pi/4$, $\bar\theta_2=0$ and $\bar\sigma_{1,\varphi} = \bar\sigma_{1,\theta} = \bar\sigma_{2,\varphi}= \bar\sigma_{2,\theta} = \pi/9$. The power angular density $f(\varphi,\theta)$ of the EMI is modeled in the same way as~\eqref{eq:non-isotropic_pdf} with fixed $\bar\varphi= -\pi/4$ and $\bar\theta=0$, while $\rho = 20$\,dB. 
Fig.~\ref{simulation-Non-Isotropic-Scattering} illustrates the SNR in~\eqref{eq:optimized_SNR} for a varying $N$, different values of $\bar\sigma_{\varphi} = \bar\sigma_{\theta}$. The SNRs with isotropic ${\bf R}={\bf R}^{\rm{iso}}$ (given by~\eqref{eq:correlation_matrix}) and without EMI are also reported. The lowest SNR is achieved with isotropic EMI. The SNR increases as $\bar\sigma_{\varphi} = \bar\sigma_{\theta}$ reduces since the overlap between the domains of $f(\varphi,\theta)$ and $f_{\textcolor{red}{1}}(\varphi,\theta)$ decreases. Compared to the case without EMI, a large gap is observed even for the relatively small angular interval $\bar\sigma_{\varphi} = \bar\sigma_{\theta} = \pi/4$, for which the domains of $f(\varphi,\theta)$ and $f_{\textcolor{red}{1}}(\varphi,\theta)$ overlaps marginally. For all the investigated scenarios, the SNR does not scale as  in the case without EMI. All this is in agreement with Proposition~\ref{prop:hardening_noise}.

\begin{remark}
The unbounded behavior of the SNR, as $N \to \infty$, is a consequence of the models in Section II, obtained under the classical far-field assumption. The latter breaks down as $N \to \infty$. To ensure the convergence of the SNR towards a finite upper limit, the models must be refined to capture the geometric near-field properties, e.g.,~\cite{Bjornson2020b} for deterministic propagation conditions. However, the importance of asymptotics is not to quantify the performance limits but rather to gain insights into any practical RIS with a large but finite $N$. This is exactly how the asymptotic findings in Proposition~\ref{prop:hardening_noise} were used in Figs.~\ref{simulation-Asymptotic-normalized} and~\ref{simulation-Non-Isotropic-Scattering}.
\end{remark}

\vspace{-0.2cm}

\section{EMI-aware RIS configuration }
So far, we have considered an RIS that is optimized against thermal noise only. It is interesting to analyze if better performance can be achieved by tuning the RIS based on the EMI statistics, i.e., knowledge of the spatial correlation matrix ${\bf R}$. To this end, we redefine the effective channel $\vect{g}_2$ in~\eqref{eq:received-signal2} as $\vect{g}_2 = \vect{H}_2\boldsymbol{\Gamma}\boldsymbol{\phi}$ where ${\bf H}_2 = \diag(h_{2,1},\ldots,h_{2,N})$ and $\boldsymbol{\phi} = [e^{\imagunit \phi_1},\ldots,e^{\imagunit \phi_N}]^{\Ttran}$. We also assume that the direct link is negligible. Hence,~\eqref{eq:SNR} takes the  form
\begin{align}\label{eq:SNR_v1}
{\mathrm {SNR }} &=\frac{P}{\sigma_w^2}\frac{|\boldsymbol{\phi}^{\Htran} \vect{a}|^2}{\boldsymbol{\phi}^{\Htran} \vect{B}\boldsymbol{\phi}+ 1}
\end{align}
with $\vect{a} = \boldsymbol{\Gamma}^{\Htran}\vect{H}_2^{\Htran}  \vect{h}_1$ and $\vect{B} =\frac{A \sigma^2}{\sigma_w^2}\boldsymbol{\Gamma}^{\Htran}\vect{H}_2^{\Htran}\vect{R}\vect{H}_2\boldsymbol{\Gamma}$. Since $\frac{1}{N}\boldsymbol{\phi}^{\Htran}\boldsymbol{\phi}=1$, we may rewrite~\eqref{eq:SNR_v1} as 
\begin{align}\label{eq:SNR_v2}
{\mathrm {SNR }} = \frac{P}{\sigma_w^2} \frac{ \boldsymbol{\phi}^{\Htran} {\bf A} \boldsymbol{\phi}}{\boldsymbol{\phi}^{\Htran}{\bf C}\boldsymbol{\phi}}
\end{align}
with ${\bf A}  = \vect{a}\vect{a}^{\Htran}$ and $\vect{C} = \vect{B} + \frac{1}{N}{\bf I}_N$. The maximization of the SNR in~\eqref{eq:SNR_v2} requires to solve the following problem:
\begin{align}
\max_{\bar{\boldsymbol{\phi}} ={\bf C}^{1/2}\boldsymbol{\phi}, |[\boldsymbol{\phi}]_n| =1, \forall n} \quad {\bar{\boldsymbol{\phi}}}^{\Htran} {\bf D} \bar{\boldsymbol{\phi}}
\end{align}
with $\bar{\boldsymbol{\phi}} ={\bf C}^{1/2}\boldsymbol{\phi}$ and ${\bf D}  = ({{\bf C}^{-1/2}})^{\Htran}{\bf A}{\bf C}^{-1/2}$. Finding its solution is hard since the problem is not convex, due to the unit-modulus constraint of the RIS, i.e., $|[\boldsymbol{\phi}]_n| =1, \forall n$. A heuristic algorithm can be designed on the basis of the projected gradient descent method that starts from some initial solution $\bar{\boldsymbol{\phi}}_0$ and then: $1$) Compute $\bar{\boldsymbol{\phi}}_{i+1} = \bar{\boldsymbol{\phi}}_{i}  + \alpha\vect{D}\bar{\boldsymbol{\phi}}_{i} $; $2$) Set $\boldsymbol{\phi}_{i+1} = e^{\imagunit \arg(\vect{C}^{-1/2}\bar {\boldsymbol{\phi}}_{i+1} )}$; $3$) Iterate until convergence. Following~\cite{Tranter2017}, we set $\alpha = \beta/\lambda_{\max}({\bf D}^{\Htran}{\bf D})$	with $\beta \in [0,1]$.
\begin{figure} 
        \centering\vspace{-5mm}
	\begin{overpic}[width=0.95\columnwidth,tics=10]{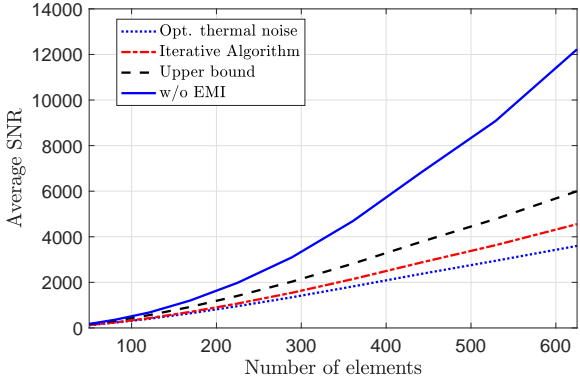}
\end{overpic} 
\vspace{-2mm}
        \caption{Average SNR  by tuning the RIS based on EMI statistics. The setup is the same of Fig.~\ref{simulation-Non-Isotropic-Scattering} with $\bar\sigma_{\varphi} = \bar\sigma_{\theta} = \pi/4$ and $\rho =15$\,dB.}
                        \label{simulation-EMI-aware} \vspace{-0.6cm}
\end{figure}

Fig.~\ref{simulation-EMI-aware} shows the average SNR achieved by the above method in the same setup of Fig.~\ref{simulation-Non-Isotropic-Scattering} with non-isotropic EMI and $\bar\sigma_{\varphi} = \bar\sigma_{\theta} = \pi/4$. We see that the iterative algorithm provides a slightly larger SNR but with approximately the same scaling behavior. If an isotropic $\bf R$ for the EMI is considered (not shown for space limitations), we notice that no gains are obtained. Fig.~\ref{simulation-EMI-aware} also shows the upper bound that is obtained by the iterative algorithm if we set $\sigma^2_w=0$ and relax the unit-modulus constraint. The scaling behavior of that upper bound is still linear, not quadratic as in the case without EMI.

\vspace{-0.3cm}
\section{Conclusions}
RIS-aided communications will always operate in the presence of EMI, unless in an anechoic chamber designed to completely absorb reflections of uncontrollable electromagnetic waves. We provided a physically meaningful model for the EMI and used it to evaluate its impact on the end-to-end SNR of an RIS-aided communication system for operating in a random scattering environment. The analysis showed that EMI may have a severe impact, especially when the number of passive elements $N$ grows large and/or when a direct link is present. While the RIS can make the received signal power grow as $N^2$, it will also reflect EMI power that is generally proportional to $N$. Hence, in the asymptotic regime as $N\to\infty$, the SNR grows as $N$. It is only in the case when the spatial correlation matrices of the EMI and the \textcolor{red}{effective} channel from the RIS to the destination are asymptotically orthogonal that the SNR will grow as $N^2$, as reported in previous literature that neglected EMI.

The analysis considered an RIS that is optimized against thermal noise only, since the EMI statistics are hard to estimate in practice. A heuristic method showed that slightly better performance can be achieved by tuning the RIS based on EMI statistics, but the SNR scaling behavior remains linear.

\vspace{-0.3cm}
\bibliographystyle{IEEEtran}

\bibliography{IEEEabrv,refs}

\end{document}